\documentstyle[12pt]{article}
\begin{document}
\baselineskip .2in
\begin{flushright}
SHEP-90/91-35 \\
August \\
\end{flushright}
\vspace{30mm}
\begin{center}
{\LARGE Non-Perturbative Two-Dimensional \\ Quantum Gravity, Again\footnote
{Talk given by SD at workshop on Random Surfaces and 2D Quantum Gravity,
Barcelona 10-14 June 1991, proceedings to appear in Nucl. Phys. {\bf B}
Proc. Suppl.}\\}
\vspace{40mm}
Simon Dalley\footnote{Address after 15th September 1991: Joseph Henry
 Laboratories, Princeton University, Princeton NJ 08544, U.S.A.} , Clifford
 Johnson and Tim Morris\\
\vspace{5mm}
{\it Department of Physics, University of Southampton, \\
Southampton SO9 5NH, U.K.}\\
\end{center}
\newpage
\noindent This is an updated review of recent work done by the authors on a
proposal for non-perturbatively stable 2D quantum gravity coupled to
$c<1$ matter, based on the flows of the (generalised) KdV
 hierarchy\cite{us1,us2}.

Since the discovery of the continuum limit of 2D quantum gravity coupled to
minimal matter \cite{gm,dooglass} to all orders in the genus expansion, the
question of a non-perturbative definition of these theories has remained.
In refs.\cite{us1,us2} we have presented a `physical'
non-perturbative definition by
elevating the fundamental (generalised) KdV flow
symmetry, known to exist perturbatively, to a non-perturbative principle.
Concentrating on KdV for simplicity, these flows are \cite{gm2};
\begin{equation}
\frac{\partial u}{\partial t_{k}} = R_{k+1}^{'}[u] = \xi_{k+1} u
\label{flow}
\end{equation}
where
\begin{equation}
\sum_{k} (k+1/2) t_{k}R_{k} - z = {\cal R} = 0 .\label{HMM}
\end{equation}
Here the string susceptibility $u=\Gamma^{''}$, prime is $d/dz$ where
 $z=\mu/\nu$, while $\mu$ and $\nu$
are the renormalised `cosmological constant' and $1/N$. The parameters
 $t_{k}/\nu$ couple to the matrix--model integrated local operators ${\cal
 O}_{k}$ on the surfaces and $R_{k}$ are the Gelfand-Dikii differentials
 \cite{gd}. The flows determine correlators of the local operators:
\begin{equation}
-\frac{\partial^{2}}{\partial z^{2}}
 < {\cal O}_{k_{1}} \ldots {\cal O}_{k_{n}}>
= \frac{\partial}{\partial t_{k_{1}}}
 \ldots \frac{\partial}{\partial t_{k_{n}}} u
\end{equation}
The string equation (\ref{HMM}) is the usual
 condition imposed by the hermitian matrix model (HMM) \cite{gm}, leading to
non-perturbative instability in particular for pure gravity
and probably all unitary $c<1$ matter. Let us relax this condition
and ask for the most general string equation compatible with the flows
 (\ref{flow}). To derive it note that dimensionally if $[u]=1$ then $[z]=-1/2$
and $[t_{k}]=-(k+1/2)$ as follows from (\ref{flow}). Then $u$ as a function
of its dimensionful arguments must have the scaling symmetry\footnote{We
assume initially that no new dimensionful parameter arises at the
 non-perturbative level.};
\begin{equation}
\sum^\infty_{k=0} (k+1/2)\  t_k
\frac{\partial u}{\partial t_k} +\frac{1}{2} z
u^{'} + u =0 \label{scale}
\end{equation}
Using (\ref{flow}) and the recurrence relation \cite{gd};
\begin{equation}
R'_{k+1} = \frac{1}{4} R'''_k - u R'_k
-\frac{1}{2} u' R_k
\end{equation}
this becomes:
\begin{equation}
\frac{1}{4} {\cal R}''' -u{\cal R}'-
\frac{1}{2}u'{\cal R} = 0 \label{difhap}
\end{equation}
Multiplying by ${\cal R}$ and integrating gives the string equation \cite{us1}:
\begin{equation}
u{\cal R}^2-\frac{1}{2} {\cal R} {\cal R}^{\prime\prime}
+\frac{1}{4} \left({\cal R}^{\prime}\right)^2 = 0 \label{smiley}
\end{equation}
The constant of integration (which is an order $\nu^{2}$ term) has been set to
 zero by hand in order that the HMM equation
${\cal R}=0$  satisfies (\ref{smiley}) asymptotically. This is  necessary
to reproduce the perturbative (genus) expansion at $z \rightarrow +\infty$.
Note that the asymptotic expansion of the solution to (\ref{smiley}) is
uniquely determined once we have fixed the spherical approximation.

As an illustrative case we can consider pure gravity $t_{m}= t_{m}\delta_{m2}$.
In the spherical approximation, where we can neglect derivatives, (\ref{smiley}
) becomes:
\begin{equation}
u(u^{2}-z)^{2} = 0 \label{sphere}
\end{equation}
As $z \rightarrow +\infty$ we take $u = \sqrt{z}$ and the corrections
will yield standard genus expansion, by construction.
If we require $u$ to have a real asymptotic expansion at $z \rightarrow
-\infty$ then we must take the root $u=0$. The susceptibility then begins
at the `torus' level and in fact has non-vanishing contributions only every 2nd
 genus.
More generally it is every $m^{\rm th}$ genus\footnote{The same asymptotic
 behaviour
has also recently been found in the unitary matrix model with `external
fields' \cite{gross}} \cite{us1}:
\begin{equation}
u = -\frac{1}{z^{2}} \sum_{n=0}^{\infty} C_{n}
 \left(\frac{1}{z^{2m+1}}\right)^{n} \;\;\; ; \;\;C_{0}=1/4 \;\; ; \;\; z
 \rightarrow -\infty
\end{equation}
It is highly likely that all other non-perturbative solutions to (\ref{smiley})
 reproducing
the standard genus expansion at $+ \infty$ are complex or have poles on the
 real axis and
are thus physically unacceptable\cite{dav}. The $z \rightarrow \pm \infty$
 behaviour
is now enough to fix completely the boundary conditions for (\ref{smiley})
 \cite{us2}
 and the (unique) numerical solution we found for pure gravity is shown in
figure 1. On the question of uniqueness we note also that at $t_{m}= t_{m}
\delta_{m1}$, (\ref{smiley}) is reducible to Painlev\'e II in $\chi$ by the
 transformation $u = 2\chi^{2} +z$. With our physical boundary conditions
it has already been shown analytically that there is a unique, real, pole-free
solution \cite{Mc}.

Thus, starting from the KdV flows and requiring physically
acceptable behaviour in choosing boundary conditions, we are
led to the solution for pure gravity shown in figure 1. Considering
$u$ as the potential for a hamiltonian operator $H$, whose eigenvalues govern
 the
non-perturbative positions of charges in a Dyson gas, we see that the spectrum
is continuous and bounded below. In fact our analysis is
equivalent to taking the continuum limit of an appropriate (critical) Dyson gas
 on the
positive real line, as we show in a moment. We can compare our stabilisation
 with results provided by
the SUSY D=1 / stochastic quantisation of the HMM investigated in
refs.\cite{stoc}. The non-perturbative difference between the two at positive
$z$ is
obvious from the manifestly different behaviour of the solutions at $z
\rightarrow -\infty$. For the model investigated in \cite{stoc}, since $u$ is
 real and $\propto
\sqrt{-z}$ at $z \rightarrow -\infty$ \cite{tim,dav1}, the susceptibility
cannot satisfy (\ref{smiley}) and so violates the KdV flows non-perturbatively.
 Also there still appear to be non-perturbative
ambiguities in the approach of \cite{stoc} in its most general form, while
 there are none for our approach.

Our analysis so far has dealt directly with the continuum limit and it is
natural to ask whether there is an appropriate formulation in terms of a
Dyson gas. The universality class of
the critical behaviour in such a gas can be conveniently characterised by the
 structure of
the eigenvalue density in the scaled neighbourhood of the end of an arc of
eigenvalues, in the  spherical approximation\cite{tomdikarry}.
 Figure 2(a) shows
this region as appropriate for pure gravity ($m=2$) at $z>0$. Eigenvalues are
concentrated on the cut (wavy line) ended by a square root branch point, and
 formally
extending the expression for the density off this cut, its first integral has
the interpretation of effective potential for one eigenvalue, $V_{\rm eff}$
 \cite{dav1}. For the $m$th critical point there are $m-1$ extra zeros of the
density in the scaled neighbourhood of the branch point, shown as dots in the
 figure. The behaviour as a function of $z$ is easily
calculated by using the WKB approximation for the spectrum
of the co-ordinate operator $H$, with the appropriate `potential' $u$
from (\ref{sphere}). This also facilitates a comparison of the extremal values
of $V_{\rm eff}$ with the leading exponential corrections to the asymptotic
solutions to (\ref{smiley}) \cite{us2}.
As $z \rightarrow 0$ the branch point
and the extra zero in figure 2(a) collide and at first it is difficult to see
 how a real, sensible
answer is maintained for $z<0$. A solution is shown in figure 2(b) whereby
the density now diverges, rather than vanishes, like a square root at the
branch point, thus generating an extra zero. This scenario is thoroughly
natural if there is a `wall' in the problem, which the branch point hits at
$z=0$. In this case the {\em two} zeros depart into the
complex plane perpendicular to the real axis as $z$ becomes more negative.
The structure shown in figure 2(a) to the left of this wall is then in the
sense of analytic continuation ($V_{\rm eff} = \infty$ strictly). Simply
speaking the wall will prevent eigenvalues leaking out of the arc and this is
in fact the origin of non-perturbative stability in our formulation, as we now
show.

We can derive the full continuum limit \`{a} la Douglas \cite{dooglass} which
also indicates the inclusion of the generalised KdV hierarchy.
Insertions of the eigenvalue operators $-\lambda$, $-d/d\lambda$ have, for
the $m$th critical point of the one-matrix model, the continuum limits:
\begin{equation}
-H = Q=d^{2} - u  \;\;\;;\;\;\;P=\sum_{i=0}^{2m-1} \alpha_{i}d^{i}
 \;\;\;;\;\;\;
d=d/dz
\end{equation}
The eigenvalue space of the HMM is ${\rm I\!R}$ and the canonical momentum
operator is $P$ (translations). The c.c.r. $[P,Q]=1$ determines the
 $\alpha_{i}:\;i<2m-1$ uniquely ($\alpha_{2m-1}$ is an undetermined
non-universal constant), the order $d^{0}$ of this relation being ${\cal
 R}^{'}=0$. Our stabilisation scheme is equivalent to the restriction to ${\rm
 I\!R}_{+}$ i.e. the imposition of a `wall' in the scaling region. To show this
we note that the canonical momentum on ${\rm I\!R}_{+}$ is $\tilde{P}$, the
generator of scale transformations, satisfying the new c.c.r.
 $[\tilde{P},Q]=Q$. If $Q$ is conjugate to macroscopic loop length
then these local scale transformations are related to physical scale
 transformations of the string/one-dimensional universe.
 We can find $\tilde{P}$
by introducing the fractional-power pseudo-differential operator and its
 differential part:
\begin{equation}
Q^{m+1/2}  =  d^{2m+1}+ (m+1/2)\{u,d^{2m-1}\} + \cdots + \{R_{m+1}[u],
d^{-1}\} \cdots
\end{equation}
\begin{equation}
[Q_{+}^{m+1/2},Q]  =  R_{m+1}^{'}
\end{equation}
We expect $\tilde{P}=\sum_{i=0}^{2m+1} \alpha_{i}d^{i}$ but $Q_{+}^{m+1/2}$
does not quite do the job. To obtain a term $d^{2}$ on the r.h.s. of the
c.c.r. we must use most generally:
\begin{equation}
\tilde{P} = \alpha_{2m+1} Q_{+}^{m+1/2} - \frac{z}{2}d
\end{equation}
giving the (differential of the) string equation;
\begin{equation}
[\tilde{P},Q] = \alpha_{2m+1}R_{m+1}^{'}  +\frac{zu^{'}}{2}  + d^{2} = d^{2}
- u. \label{PQ}
\end{equation}
This is (\ref{difhap}) at the $m$th critical point, modulo  $\alpha_{2m+1}$
which may be absorbed into the string coupling.
By explicitly taking the continuum limit of the Dyson gas on ${\rm I\!R}_{+}$,
realised for example by the complex matrix model \cite{us1}, one finds that the
 constant arising from an integration of (\ref{PQ}) is zero as in
 (\ref{smiley}). Moreover the physical
boundary conditions we chose earlier are precisely the ones appropriate to this
 gas.
In the spherical approximation $u$ is the scaling part of the end of the
eigenvalue density. For $z>0$ the `wall' has no affect on the WKB
expansion, in particular the leading term is $\sqrt{z}$ as in the HMM. At
$z<0$ the end of the density remains fixed at the wall so $u=0$ in the
spherical approximation.

It is natural to suppose that these considerations have analogues for the
generalised KdV hierarchy, related to the full set of $(p,q)$ minimal models
\cite{dooglass,kut}. One would generalise to:
\begin{equation}
Q= d^{q} +\sum_{i=0}^{q-2}u_{i}d^{i} \;\;\;\;\;\; P= Q_{+}^{p/q}
\end{equation}
\begin{equation}
\tilde{P}= Q_{+}^{1+p/q} - \frac{z}{q}d
\end{equation}
Some details of this will be explored in \cite{bill}.

Let us finally consider  the Dyson-Schwinger  equations\footnote{Some of what
 follows was elaborated while these notes were being prepared.} of the
one-matrix models, at least that part of them which may be represented as
formal Virasoro constraints \cite{Vir}. Our differentiated string equation
(\ref{smiley}) follows from the constraint $L_{0}\tau=0$
(where $u= -2d^{2}\log \tau$ and $z \rightarrow z+t_{0}$) on using the KdV flow
 (\ref{flow}). This is not suprising in that it is precisely
the expression of scale invariance. The higher Virasoro constraints:
\begin{equation}
L_{n}\tau  =  0\;\;:\;\;n\geq1
\end{equation}
where
\begin{equation}
L_{n} = \sum_{k=0}^{\infty} \left( k+\frac{1}{2} \right) t_{k} \frac{\partial}{
\partial t_{k+n}} + \frac{1}{4} \sum_{k=1}^{n} \frac{\partial^{2}}{
\partial t_{k-1} \partial t_{n-k}}
\end{equation}
follow by applying the recursion operator which generates
higher symmetries of the KdV hierarchy \cite{book}\footnote{H.Kawai informed me
 that the DS constraints $\{L_{n}\tau=0:n\geq 0\}$
should in turn imply that $\tau$ is a $\tau$-function of the KdV
hierarchy subject to $L_{0}\tau=0$
 (see his contribution to these proceedings).}.
The HMM string equation derives from $L_{-1}\tau=0$ and one might ask what
 happens to this constraint in our formulation.
 Since we are working on ${\rm I\!R}_{+}$ there is no translation invariance
 and the $L_{-1}$ constraint
is naively absent. More precisely the relevant DS equation picks up a boundary
 term from the
`wall':
\begin{equation}
L_{-1}\tau = \frac{\partial \tau}{\partial \sigma} \label{lostsheep}
\end{equation}
where $\sigma$ is the scaled position of the `wall', which up to now we have
taken as the origin. In fact there is
no reason why we should choose $\sigma=0$ and more correctly we might consider
 it
as an extra (non-perturbative) parameter $u=u(z,t_{k},\sigma)$. The correct
canonical momentum gets shifted now to $\tilde{P}_{\sigma}=\tilde{P} +
 \sigma P$ and thus
\begin{equation}
[\tilde{P}_{\sigma}, Q+\sigma]  =  Q +\sigma
\end{equation}
implying
\begin{equation}
\frac{1}{4} {\cal R}''' -u{\cal R}'-
\frac{1}{2}u'{\cal R} + \sigma{\cal R}' = 0 \label{new}
\end{equation}
which may now be integrated to give the string equation.
But $\sigma$, which has the same dimension as $u$, now contributes to the
 scaling
equation (\ref{scale}) a term $\sigma \partial u/\partial \sigma$. Comparing
 with
(\ref{new}) we identify
\begin{equation}
\frac{\partial u}{\partial \sigma} = {\cal R}'
\end{equation}
which is equivalent to (\ref{lostsheep}). The other Virasoro
constraints also pick up boundary terms at general $\sigma$
\begin{equation}
L_{n}\tau = \sigma^{n+1}\frac{\partial \tau}{\partial \sigma}
\end{equation}
as follows simply from varying the boundary of the eigenvalue integration as
$\sigma \rightarrow \sigma +\epsilon \sigma^{n+1}$. The parameter $\sigma$ can
 be set to zero by an analytic redefinition of the $t_{k}$ together with
 $u-\sigma \rightarrow u$. In this sense
it is redundant and in fact if we assume, following the HMM, that the
 non-perturbative
loop expectation is $\Psi(l) = < {\rm e}^{-lH} >$, $\sigma$ corresponds to a
 boundary
cosmological constant \cite{moore}. The loop wavefunction in the scaling limit
 is
\begin{equation}
\Psi(l) \propto  \int_{\sigma}^{\infty} \rho(\psi){\rm e}^{-l\psi} d\psi
\end{equation}
where
\begin{equation}
\rho(\psi) \propto \int_{\mu}^{\infty} <x \mid \delta (\psi -H) \mid x> dx
\end{equation}
Hence for $\sigma>0$ the wavefunction is guaranteed to have sensible
exponentially decreasing large $l$ behaviour.
\vspace{5mm}

\noindent {\bf Acknowledgements:} S.D. would like to thank the organisers for
 the opportunity
to present this work and workshop participants for their interest. Financial
support from the S.E.R.C. is acknowledged by S.D. and C.J.


\newpage
\begin{thebibliography}{99}
\bibitem{us1} S.Dalley, C.Johnson and T.Morris, Southampton Univ. preprint
SHEP 90/91-16.
\bibitem{us2} S.Dalley, C.Johnson and T.Morris, Southampton Univ. preprint
SHEP 90/91-28.
\bibitem{gm} D.J.Gross and A.A.Migdal, Phys. Rev. Lett.{\bf 64} (1990) 127;
 M.R.Douglas and S.H.Shenker, Nucl. Phys. {\bf B335} (1990) 135;
E.Br\'ezin and V.A.Kazakov, Phys. Lett. {\bf B236} (1990) 144.
\bibitem{dooglass} M.R.Douglas, Phys. Lett. {\bf B238} (1990) 176.
\bibitem{gm2} D.J.Gross and A.A.Migdal, Nucl. Phys. {\bf B340} (1990) 333;
 T.Banks, N.Seiberg, M.R.Douglas and S.H.Shenker, Phys. Lett.
{\bf B238} (1990) 279.
\bibitem{gd} I.R.Gelfand and L.A.Dikii, Russian Math. Surveys {\bf 30} (1975)
77.
\bibitem{dav} F.David, Mod. Phys. Lett. {\bf A5} (1990) 1019.
\bibitem{stoc} E.Marinari and G.Parisi, Phys. Lett. {\bf B240} (1990) 375;
 J.Ambj{\o}rn, J.Greensite and S.Varsted, Phys. Lett. {\bf B249}
(1990) 411; E.Marinari and G.Parisi, Phys. Lett. {\bf B247} (1990) 537;
M.Karliner and A.A.Migdal, Mod. Phys. Lett. {\bf A5} (1990) 2565;
J.Ambj{\o}rn and J.Greensite, Phys. Lett. {\bf B254} (1991) 66.
\bibitem{gross} D.J.Gross and M.J.Newman, Princeton preprint PUPT-1257.
\bibitem{Mc} S.P.Hastings and J.B.Macleod, Arch. Rat. Mech. and Anal.
{\bf 73} (1980) 31.
\bibitem{tim} J.Ambj{\o}rn, C.Johnson and T.Morris, preprint SHEP 90/91-29 /
 NBI-HE-91-27.
\bibitem{dav1} F.David, In Proceedings of workshop on Random Surfaces,
Quantum Gravity and Strings, Carg\`ese, May 28-June 1 1990.
\bibitem{tomdikarry} S.Dalley, Phys. Lett. {\bf B253} (1991) 292; P.Ginsparg
 and J.Zinn-Justin, Phys. Lett. {\bf B255} (1991) 189; H.Neuberger, Nucl. Phys.
 {\bf B352} (1991) 689; F.David, Nucl. Phys. {\bf B348} (1991) 507.
\bibitem{kut} P. di Francesco and D.Kutasov, Nucl. Phys. {\bf B342} (1990) 589.
\bibitem{bill} C.Johnson, T.Morris and B.Spence, in preparation.
\bibitem{Vir} M.Fukuma, H.Kawai and R.Nakayama, Int. J. Mod. Phys. {\bf A6}
(1991) 1385; R.Dijkgraaf, E.Verlinde and H.Verlinde, Nucl. Phys. {\bf B348}
(1991) 435.
\bibitem{book} P.J.Olver, {\em Applications of Lie Groups to
Differential Equations}, Springer-Verlag, 1986.
\bibitem{moore} E.Martinec, G.Moore and N.Seiberg, preprint RU-14-91 /
 YCTP-P10-91 / EFI-91-14.
\end{thebibliography}
\end{document}